\documentclass[12pt,letterpaper]{article} 

\usepackage[includeheadfoot,
            marginratio={1:1,2:3}, 
            width=412pt, 
            height=688pt,]{geometry}

\usepackage{amsmath}
\usepackage{amsfonts}
\usepackage{amssymb}

\newcommand{\eq}[1]{\begin{equation}
                     \begin{split} #1 \end{split}
                     \end{equation}}
\newcommand{\fa}{\hat}

\newcommand{\Lie}{{\cal L}} 

\begin{document}

\vspace*{-1.5cm}
\begin{flushright}
  {\small
  MPP-2012-133\\
  DFPD-2012-TH-12

  }
\end{flushright}

\vspace{1.5cm}
\begin{center}
{\LARGE
A bi-invariant 
  Einstein-Hilbert  action \\[0.2cm]
for the non-geometric string
}
\vspace{0.2cm}

\end{center}

\vspace{0.35cm}
\begin{center}
  Ralph Blumenhagen$^{1}$, Andreas Deser$^{1}$, Erik Plauschinn$^{2,3}$ \\ and
Felix Rennecke$^{1}$
\end{center}

\vspace{0.1cm}
\begin{center}
\emph{$^{1}$ Max-Planck-Institut f\"ur Physik (Werner-Heisenberg-Institut) \\
   F\"ohringer Ring 6,  80805 M\"unchen, Germany } \\[0.1cm]
\vspace{0.25cm}
\emph{$^{2}$ Dipartimento di Fisica e Astronomia ``Galileo Galilei'' \\
Universit\`a  di Padova, Via Marzolo 8, 35131 Padova, Italy}  \\[0.1cm]
\vspace{0.25cm}
\emph{$^{3}$ INFN, Sezione di Padova \\
Via Marzolo 8, 35131 Padova, Italy}  \\

\end{center}

\vspace{1cm}


\begin{abstract}
Inspired
by recent studies on string theory with  non-geometric fluxes,  
we develop a differential geometry calculus combining
usual diffeomorphisms with what we call  $\beta$-diffeomorphisms.
This allows us  to construct a manifestly  
bi-invariant Einstein-Hilbert type action
for the graviton, the dilaton  and a  dynamical (quasi-)symplectic structure.
The equations of motion of this symplectic gravity theory, 
further generalizations and
the relation to the usual form of the string effective action are discussed.
The Seiberg-Witten limit, known for open strings to  relate commutative with
non-commutative theories, makes an interesting appearance.   
\end{abstract}

\clearpage

\section{Introduction}
\label{sec:intro}

String theory is expected to be a consistent theory of
quantum gravity. In this respect, it is interesting to note that
a generic feature of all known string
theories is that besides the graviton, there exist
two additional massless excitations, 
the  Kalb-Ramond field $B_{\mu\nu}$ and the dilaton $\phi$.
At leading order, the dynamics of these fields is governed 
by the extension of the Einstein-Hilbert action
\eq{
\label{stringaction}
S=-{1\over 2\kappa^2}
\int 
\hspace{-0.75pt}
d^nx \hspace{1pt}\sqrt{-G}\hspace{1pt} e^{-2\phi}\Bigl(R-{\textstyle{1\over 12}} \hspace{0.5pt} H^2
+4 \hspace{0.25pt} (\partial \phi)^2
 \Bigr) \hspace{0.5pt},
}
which has two types of local symmetries. Namely, it is invariant 
under diffeomorphisms of the space-time
coordinates,  and under gauge transformation of the Kalb-Ramond field.
Note also, this action is a valid approximation for solutions with large radii.

Employing T-duality \cite{Shelton:2005cf},
methods of generalized geometry \cite{Grana:2008yw,Coimbra:2011nw,Berman:2012vc}
and double field theory \cite{Siegel:1993th,Hull:2009mi,Hohm:2010jy,Aldazabal:2011nj}, 
it has become clear 
during the last years
that also
a  non-geometric frame exists, where the degrees of
freedom are described by a metric on the co-tangent bundle,
by a dilaton  and by a (quasi-)symplectic structure $\beta^{ab}$. 
The latter gives rise to 
so-called non-geometric $Q$- and $R$-fluxes. In particular,
the $R$-flux has been argued to be related to a
non-associative  structure 
\cite{Bouwknegt:2004ap,Blumenhagen:2010hj,Lust:2010iy,Blumenhagen:2011ph,Mylonas:2012pg}. 
However, in contrast to the well-established
non-commutative behavior of open strings \cite{Seiberg:1999vs}, the
generalization to closed strings is  more complex,
as in a gravitational theory 
the non-commutativity parameter is expected to be dynamical.

Since in the non-geometric frame, apart from the dilaton,  
one deals with just a  metric and a (quasi-)symplectic structure,
it is natural to expect that both local symmetries
of the string action can be given a description in terms
of a (generalized) differential geometry. Starting
from  so-called double field theory, this question
has already been approached
in an interesting way
in
\cite{Andriot:2012wx,Andriot:2012an} (see also \cite{Andriot:2011uh}).
However, the action derived in \cite{Andriot:2012wx,Andriot:2012an}
is not  manifestly
invariant under both local symmetries.
It is the objective  of this letter
to construct such a manifestly bi-invariant action for the non-geometric
string. The appropriate  mathematical framework for this turns out to be
the theory of Lie and Courant algebroids 
\cite{Halmagyi:2008dr, Blumenhagen:2012pc}, which we
will mention only briefly.
More details on the underlying mathematical structure
of Lie algebroids 
and the details of the  computations will appear in 
\cite{BDPR2012}.

Here, we  present the main steps of a  construction
of an Einstein-Hilbert type action, which is manifestly
invariant under both usual diffeomorphisms and 
what we call $\beta$-diffeomorphisms.
This bi-invariant action  turns out to be closely related to the  action
derived for  non-geometric fluxes using double field 
theory \cite{Andriot:2012wx,Andriot:2012an}. 
Remarkably, relations familiar 
from  the Seiberg-Witten limit for D-branes in a two-form background
also appear  in this closed-string framework.

 
\section{$\beta$-diffeomorphisms}

As mentioned in the introduction,
in addition to the dilaton, we consider
the co-tangent bundle T$^\star$M of a  manifold with metric 
$\fa g= \fa g^{ab}\, e_a\otimes e_b$ and
an invertible anti-symmetric bi-vector 
$\fa\beta ={1\over 2} \fa\beta^{ab} e_a \wedge e_b=\fa\beta^{ab} e_a \otimes
e_b$, where our notation is $e_a=\partial_a$ and
$e^a=dx^a$. Note that  $\fa\beta$ can
be thought of as a \mbox{(quasi-)}symplectic structure giving rise
to a \mbox{(quasi-)}Poisson structure $\{f,g\}=\fa\beta^{ab}\,\partial_a f\, \partial_b g$,
with  Jacobi-identity  
${\rm Jac}(f,g,h)=\fa\Theta^{abc}\, \partial_a f\, \partial_b g\, \partial_c h$.
The $R$-flux is defined as 
$ \fa\Theta^{abc} =3\,\fa\beta^{[a|m}\partial_m\fa\beta^{|bc]}$,
where the \mbox{(anti-)}symmetrization of indices contains a factor of $(1/n!)$. 
Moreover,  $\fa\beta$ provides a natural (anchor)
map $\beta^\sharp :{\rm T}^\star {\rm M}\to {\rm TM}$
via $\beta^\sharp e^a=\hat\beta^{am} e_m$. As we will see, it is essential
that $\fa\beta$ is invertible, which is however the generic situation.
On the other hand,
that means we can only describe backgrounds for which that requirement is satisfied.

Compared to the standard differential geometry calculus, here, not only the tangent
bundle but also  the co-tangent bundle  plays an  important
role.
This suggests that  the former principle of diffeomorphism
covariance of gravity, the equivalence principle,  
should be extended by a second class of diffeomorphisms.
Recall, that in the former case, infinitesimal
diffeomorphisms $x^a\to x^a+\xi^a(x)$ are given by
the Lie derivative $\delta_\xi X=L_\xi X$, which
acts as  the Lie bracket on vector fields and as the
anti-commutator of the insertion map and the exterior differential on forms.
For the second class, that is infinitesimal transformations 
parametrized by the components of a one-form $\fa\xi=\hat \xi_a dx^a$, we note the following.
The bracket, generalizing the commutator of vector fields
to forms, is the so-called Koszul-bracket defined as
\eq{\label{kos}
	\bigl[ \hspace{0.5pt} \fa\xi,\eta \bigr]_{\rm K}=L_{\beta^\sharp\fa\xi}\,\eta-\iota_{\beta^\sharp\eta}d\fa\xi\, ,
}
where $\iota$ denotes the insertion map.
In addition, let us define the action of a one-form on a function $\phi$ 
by the anchor map:
\eq{
  dx^a (\phi) := \beta^{\sharp}(dx^a)(\phi)= \fa\beta^{am}\partial_m \phi =:
   D^a \phi\, .
}

Now, we can proceed as in ordinary differential geometry and define tensors
by their infinitesimal transformation properties.
In particular, a scalar field $\phi$ is called a $\beta$-scalar if it  transforms as
\eq{
   \fa\delta_{\fa\xi} \phi= \mathcal{L}_{\fa\xi} \phi=\fa\xi(\phi)= \fa\xi_m\,
   D^m\phi\, ,
}
and a one-form $\eta$ is a $\beta$-one-form if
\eq{
	\fa\delta_{\fa\xi} \eta &=
        \mathcal{L}_{\fa\xi}\eta=\bigl[\fa\xi,\eta\bigr]_{\rm K} \\
	&= \left(\fa\xi_m D^m\eta_a - \eta_mD^m\fa\xi_a+\fa\xi_m\eta_n\,\fa Q_a{}^{mn}\right)e^a \,,
}
with $\fa Q_c{}^{ab} = \partial_c\fa\beta^{ab}$.
The transformation properties of general $\beta$-tensors are 
then determined by requiring  the Leibniz rule
of $\delta_{\fa\xi}$ for tensor products and contractions, which
 implies for instance that  a $\beta$-vector field $X=X^a e_a$ transforms as
\eq{ \label{vectorvar}
     	\fa\delta_{\fa\xi} X&=  \mathcal{L}_{\fa\xi} X\\
    &=\left(\fa\xi_mD^mX^a+X^mD^a\fa\xi_m-X^m\fa\xi_n\,\fa Q_m{}^{na}\right)e_a\,.
}

To continue, we  have to fix  the nature of the 
metric $\fa g^{ab}$ and the anti-symmetric bi-vector $\fa\beta^{ab}$.  The former  should be  
a tensor with respect to both diffeomorphisms and $\beta$-diffeomorphisms,
while we require the latter  only to be a tensor under diffeomorphisms.
As will become clear  below, it should  transform 
under $\beta$-diffeomorphisms non-covariantly
\eq{
  \label{betarule}
 \fa\delta_{\fa\xi} \fa \beta :=&\hspace{2pt} \Lie_{\fa\xi} \fa \beta + \fa\beta^{am}\fa\beta^{bn} \left(\partial_m \fa\xi_n -\partial_n \fa\xi_m \right) e_a \otimes e_b \\
=& \hspace{2pt}\fa\xi_m\, \fa\Theta^{mab}\, e_a\otimes e_b\, .
}
Moreover, the variation with respect to  $\fa\xi$ should commute 
with partial derivatives, i.e. $[\fa\delta_{\fa\xi},\partial_a] = 0$. 
The Lie brackets of infinitesimal diffeomorphisms and
$\beta$-diffeomorphisms are
\eq{
    \bigl[\delta_{\xi_1},\delta_{\xi_2}\bigr]&=\delta_{[\xi_1,\xi_2]}\, , \\
     \bigl[\hspace{0.5pt}\fa\delta_{\fa\xi},\delta_{\eta} \bigr]&=
      \delta_{{\cal L}_{\fa\xi}\eta}\, ,  \\
    \bigl[\hspace{0.5pt}\fa\delta_{\fa\xi_1},\fa\delta_{\fa\xi_2}\bigr]&=\fa\delta_{[\fa\xi_1,\fa\xi_2]_{\rm
        K}}   +\delta_{(\iota_{\fa\xi_1} \iota_{\fa\xi_2} \fa\Theta)} \, .
}

Ordinary differential geometry is based on the covariantization
of the partial derivative of tensors,  however, because of
\eq{
    \fa\delta_{\fa\xi} (\partial_a \phi)=\Lie_{\fa\xi} (\partial_a \phi)
      + (D^m\phi)(\partial_a \fa\xi_m-\partial_m \fa\xi_a)\, ,
}
under a $\beta$-diffeomorphism
the partial derivative of a scalar does not transform as a $\beta$-vector.
But, on the other hand, we have defined the transformation of $\fa\beta$ in
eq.~\eqref{betarule} such that the derivative
$D^a\phi$ transforms precisely as a $\beta$-vector, i.e.
$\fa\delta_{\fa\xi} (D^a\phi)=\Lie_{\fa\xi} (D^a \phi)$.
Finally, using one of the Bianchi identities derived in 
\cite{Blumenhagen:2012pc,Andriot:2012wx}, we find
that the $R$-flux is  also a $\beta$-tensor, that is 
$\fa\delta_{\fa\xi}\fa\Theta^{abc} = \Lie_{\fa\xi} \fa\Theta^{abc}$.


\section{Covariant derivative, torsion and curvature}

As established in the last section, the role played by $\partial_a$
in usual gravity theories is now taken by the derivative $D^a$.
Following the same steps as in standard differential geometry, we
then  define the covariantization of $D^a$ as
\eq{
      \hat\nabla^a X^b=D^a X^b - \hat\Gamma_c{}^{ab}\, X^c \;,
} 
and  the action on forms reads 
$\hat\nabla^a \eta_b=D^a \eta_b + \hat\Gamma_b{}^{ac}\, \eta_c$.
Demanding that the covariant derivative
is a $\beta$-tensor requires that the $\beta$-connection
cancels  the  anomalous transformation of the first term,
leading to
\eq{ \label{anomalous}
  \fa\Delta_{\fa\xi} \bigl(\hat\Gamma_c{}^{ab}\bigr) = D^a\bigl( D^b \fa\xi_c - \fa\xi_m \fa Q_c{}^{mb}\bigr) \, ,
}
with $\fa\Delta_{\fa\xi}=\delta_{\fa\xi}-\Lie_{\fa\xi}$.
Under usual diffeomorphisms, $\hat\Gamma_c{}^{ab}$  needs to transform
anomalously as
\eq{ \label{anomalousb}
  \Delta_{\xi}(\hat\Gamma_c{}^{ab}) = -D^a\left( \partial_c \xi^b \right) \, .
}

Taking the commutator of two covariant derivatives defines the $\beta$-torsion 
\eq{
    \bigl[\hat\nabla^a,\hat\nabla^b\bigr]\phi=- \hat T_c{}^{ab}\, D^c \phi \,,
}
which can be expressed as
\eq{
      \hat T_c{}^{ab}=\fa\Gamma_c{}^{ab}-\fa\Gamma_c{}^{ba}-\fa{\cal Q}_c{}^{ab} \, ,
}
with $\fa{\cal Q}_c{}^{ab}=\fa{Q}_c{}^{ab}+ \fa\Theta^{abm} \fa\beta_{mc}$.
By construction, this is a usual tensor and a $\beta$-tensor.
The curvature is defined by
\eq{
    \bigl[\hat\nabla^a,\hat\nabla^b\bigr]\, X^c= -\hat R_m{}^{cab}\, X^m 
      - \hat T_m{}^{ab}\, \hat\nabla^m X^c\, ,
}
leading to
\eq{
    \hat R_m{}^{cab} =D^a \hat\Gamma_m{}^{bc}&-D^b \hat\Gamma_m{}^{ac}
         +\hat\Gamma_n{}^{bc}\, \hat\Gamma_m{}^{an} 
    -  \hat\Gamma_n{}^{ac}\, \hat\Gamma_m{}^{bn}
                          -\fa{\cal Q}_n{}^{ab}\, \hat\Gamma_m{}^{nc}\, .
}
The metric-compatible and torsion-free Levi-Civita connection
  takes the form
\eq{
\label{LCconnect}
    \hat\Gamma_c{}^{ab}=\tilde\Gamma_c{}^{ab}-\hat g_{cq} \hspace{0.5pt}
     \hat g^{p(a|} \fa{\cal Q}_p{}^{|b)q}  
+{1\over 2} \fa{\cal Q}_c{}^{ab}\, ,
}
with
\eq{
         \tilde\Gamma_c{}^{ab}={1\over 2} \hspace{0.5pt} \hat g_{cp}\left(
                D^a \hat g^{bp} + D^b \hat g^{ap} - D^p \hat g^{ab}\right)\, .
}
Note that one can check explicitly that \eqref{LCconnect}
has the right anomalous transformation behavior under diffeomorphisms
\eqref{anomalousb} and $\beta$-diffeomorphisms \eqref{anomalous}.

For vanishing torsion, the  Ricci tensor
 $\hat R^{ab}=\hat R_m{}^{amb}$ is symmetric and reads
\eq{
      \hat R^{ab}=D^m \hat\Gamma_m{}^{ba}-D^b \hat\Gamma_m{}^{ma}
            &+\hat\Gamma_n{}^{ba}\, \hat\Gamma_m{}^{mn} 
              - \hat\Gamma_n{}^{ma}\, \hat\Gamma_m{}^{nb}\, .
}
The Ricci scalar $\hat R=\hat g_{ab} \hat R^{ab}$ can be expanded as
\eq{
\label{Rexpand}
     \hat R=-\Bigl[ \hspace{14pt} & D^a D^b \hat g_{ab} 
          - D^a\left( \hat g_{ab} \, \hat g^{mn}\,  D^b \hat g_{mn}  \right)\\[0.1cm]
          -\hspace{1.5pt}&{1\over 4} \hat g_{ab}\Bigl( D^a  \hat g_{mn}\, D^b  \hat g^{mn}
                            -2 D^a  \hat g_{mn}\, D^m  \hat g^{nb} -  \hat g_{mn}\,  \hat g_{pq}\, D^a  \hat g^{mn}\, D^b  \hat
            g^{pq}\Bigr)\\[0.1cm]
      +\hspace{1.5pt}&{1\over 4}\,  \hat g_{ab}\,  \hat g_{mn} \, \hat g^{pq}\, \fa{\cal Q}_p{}^{ma} \fa{\cal Q}_q{}^{nb}
      +{1\over 2}  \hat g_{ab}\, \fa{\cal Q}_m{}^{nb}\, \fa{\cal Q}_n{}^{ma}
     +  \hat g_{ab}\, \fa{\cal Q}_m{}^{ma}\, \fa{\cal Q}_n{}^{nb}  \\
      +\hspace{1.5pt}&2   D^a \left( \hat g_{ab} \, \fa{\cal Q}_m{}^{mb} \right)
      -  \hat g_{ab}\, \hat g_{mn}\,  D^a \hat g^{pn}  \,   \fa{\cal Q}_p{}^{bm}
     +   \hat g_{ab} \,\hat g^{mn}\,  D^a \hat g_{mn}   \,  \fa{\cal
       Q}_p{}^{bp}\hspace{10pt}\Bigr] ,
}
which is the same  expression  as  in \cite{Andriot:2012an} 
if one substitutes  $\fa{\cal Q}_a{}^{bc}\leftrightarrow {Q}_a{}^{bc}$.


\section{Bi-invariant action}

After having defined a covariant   curvature, we can now move forward
and construct a bi-invariant action for the fields
$(\hat g,\hat\beta,\phi)$, where the dilaton $\phi$  
is chosen to be a scalar under both transformations.
Since by construction $\fa\Theta^{abc}$ is  a  tensor,
the following combination
\eq{
   \hat{\cal L}=e^{-2\phi} \left(   \hat R -{1\over 12} \fa\Theta^{abc}\, \fa\Theta_{abc}
              +4\hspace{0.5pt} \hat g_{ab}\, D^a\phi D^b \phi\right)
}        
behaves as a scalar under both types  of diffeomorphisms.
Our aim is now to  construct a bi-invariant action
\eq{
     \hat S=-{1\over 2\kappa^2} \int d^nx\, \mu(\fa g, \fa \beta)\, \hat{\cal L}\, ,
}
where $\mu$ denotes an appropriate measure.

An obvious first choice would be  $\mu=\sqrt{-\fa g}$, 
however, using that $\fa g^{ab}$ is a $\beta$-tensor we find
\eq{
     \fa\delta_{\hat\xi} (\sqrt{-\fa g}\, \fa {\cal L})=
       \partial_k\left( \sqrt{-\fa g}\, \fa {\cal L}\, \fa
       \xi_m\,\right)\fa\beta^{mk} 
      -  \sqrt{-\fa g}\, \fa {\cal L}\, \fa \xi_m\, \partial_k
       \fa\beta^{mk} &\, ,
}
so that the sign in front of the last term does not complete
the desired total derivative. Furthermore, one can show that
the resulting action is not invariant
under usual diffeomorphisms either.
But because of the relation $\delta_{\hat\xi} |\fa \beta^{-1}| = 2 |\fa \beta^{-1}| \fa\xi_m\,
\partial_k \fa\beta^{mk} + \fa\xi_m \fa\beta^{mk} \partial_k |\fa \beta^{-1}|$ for the absolute value of 
\raisebox{0pt}[\ht\strutbox][0pt]{$\det( \fa\beta^{-1})$},
the missing terms can be accounted for by modifying the
measure to $\mu=\sqrt{-\fa g}\, |\fa \beta^{-1}|$.
Analogously, this new measure also ensures the diffeomorphism invariance
of the action. Note that $|\fa \beta^{-1}|={\rm Pf}(\beta^{-1})^2\ge 0$.
Thus, we have succeeded in constructing the bi-invariant action
\eq{
\label{finalaction}
      \hat S=-{1\over 2\kappa^2} &\int d^nx\, \sqrt{-\hat g}\: \bigl|\hat \beta^{-1}\bigr|\: e^{-2\phi}
     \,  \Bigl(   \hat R -{1\over 12} \fa\Theta^{abc}\, \fa\Theta_{abc}
              +4\hspace{1pt} \hat g_{ab}\, D^a\phi D^b \phi\Bigr)\, ,
} 
whose form closely resembles the universal part of the 
low-energy effective action of string theory \eqref{stringaction}.
We call this theory  {\it symplectic gravity}
with a dilaton.

The equations of motion are derived by varying the action
\eqref{finalaction} with respect to the metric, the
(\mbox{quasi-)}symplectic structure and the dilaton.
Since $\fa\beta^{ab}$ appears also through the $D^a$ derivative, this is
a non-trivial computation. 
Employing the relation
\eq{
\sqrt{-\hat g}\: \bigl|\hat \beta^{-1}\bigr| \: \hat\nabla^m X_m
=\partial_k \Bigl( \sqrt{-\hat g}\: \bigl|\hat \beta^{-1}\bigr|\, \fa\beta^{mk}\,
X_m\Bigr)\, ,
}
we  obtain the three independent equations        
\eq{
  \label{eom}
  &0= \fa R^{ab} + 2 \fa\nabla^a \fa\nabla^b \phi - \tfrac{1}{4} \fa\Theta^{amn} \fa\Theta^b{}_{mn} \,, \\
  &0 = -\tfrac{1}{2} \fa\nabla_m \fa\nabla^m \phi + (\fa\nabla_m \phi)(\fa\nabla^m \phi) - \tfrac{1}{24} 
     \fa\Theta^{mnr} \fa\Theta_{mnr} \,, \\[0.5mm]
   &0= -\tfrac{1}{2} \fa\nabla^m \fa\Theta_m{}^{ab} + (\fa\nabla^m \phi)\fa\Theta_m{}^{ab} \,.
}
These feature  the same form as the usual string-frame equations of motion
derived from the action \eqref{stringaction}.
A more detailed derivation will be presented in \cite{BDPR2012}.

Finally, a natural guess for the action of the massless bosonic states
in the Ramond-Ramond sector is
\eq{
\label{RRaction}
    \fa {\cal L}^{RR}= -\sum_n  {1\over 2\, n!}\; \fa  g_{a_1 b_1}\ldots
       \fa  g_{a_n b_n}\; \fa F^{a_1\ldots a_n}\,  \fa F^{b_1\ldots b_n}  \, ,
} 
where  $\fa F^{a_1\ldots a_n} =  n\hspace{1pt} \fa\nabla^{[a_1} \fa C^{a_2\ldots a_n]}+\mathcal O(\fa\Theta)$ and 
$n$ is even (odd)  for type IIA(B) theories.


\section{Relations to string theory}

Generalized geometry and double field
theory suggest that
the relation between the geometric and  non-geometric
fields is given by
\eq{
  \label{relation_dft}
           \tilde g &=\hphantom{-}(G+ B)^{-1}\, G\,  (G-B)^{-1} \;, \\
\tilde\beta &=- (G+B)^{-1}\, B\,  (G- B)^{-1}\;.
} 
However, 
starting from  the action \eqref{stringaction} and 
inserting this transformation,
the computation in \cite{Andriot:2012wx,Andriot:2012an} shows that 
one does not find eq.~\eqref{finalaction}. 
But,  observing that the relation between the fields 
$(G,B)$ and $(\tilde g,\tilde\beta)$ is formally the same
as the one appearing in the study of D-branes
in a two-form flux backgrounds, a second natural possibility 
arises.
In particular,
in the Seiberg-Witten limit \cite{Seiberg:1999vs}, i.e. where
a fluxed brane theory is effectively described
by a non-commutative gauge theory, the relation between
the two sets of fields reads
\eq{
\label{redefine}
    B=\hat\beta^{-1}  \, ,\qquad
     G = -\hat\beta^{-1}\, \hat g\, \hat\beta^{-1} \; .
}   
(Note that we are not taking a true limit  $G\to 0$, that is 
we are not neglecting  any terms from the action.) 
Now,
a straightforward though tedious computation to be presented
in detail in \cite{BDPR2012} shows
that indeed the two actions \eqref{stringaction} and \eqref{finalaction}
are related via  this field redefinition, i.e.
\eq{
S\bigl(G(\fa g,\fa\beta),B(\fa g,\fa\beta),\phi\bigr)=\hat S\bigl(\fa g,\fa\beta,\phi\bigr)\, .
}
As an immediate  consequence, the action which appeared
in \cite{Andriot:2012wx,Andriot:2012an}
and eq.~\eqref{finalaction}
are related via the field redefinition
$\hat\beta =\tilde\beta - \tilde g\, \tilde\beta^{-1} \tilde g$
and $\hat g = \tilde g - \tilde g\, \tilde\beta^{-1} \tilde g\,
\tilde\beta^{-1} \tilde g$.

Let us provide more arguments for the relation
among the actions.
Instead of the infinitesimal variations $\delta_{\xi}$ and
\raisebox{0pt}[\ht\strutbox][0pt]{$\fa\delta_{\hat \xi}$}, consider $\delta_{\xi}$ and
the  linear combination
\raisebox{0pt}[\ht\strutbox][0pt]{$ \check\delta_{\hat \xi} X=L_{(\beta^\sharp\hat\xi )} X -  \fa\delta_{\hat\xi} X$},
where
$X$ is assumed to be tensor 
with respect to diffeomorphisms
but not necessarily with respect to $\beta$-diffeomorphisms.
We then find
\eq{
      \check\delta_{\hat\xi} \hat g^{ab}&= 
            -2\,\hat\beta^{(a|m}\, (\partial_m \hat\xi_n-\partial_n \hat\xi_m
      )\,  \hat g^{n|b)} \, ,\\ 
     \check\delta_{\hat \xi} \hat \beta^{ab}&= 
              -\hat \beta^{am} (\partial_m \hat\xi_n-\partial_n \hat\xi_m
      )\,  \hat \beta^{nb}\, .
}
Using
then
the transformation \eqref{redefine}, we can compute the resulting
infinitesimal transformations
\raisebox{0mm}[0mm][0pt]{$\check\delta_{\hat \xi} G_{ab}=0$}
and  \raisebox{0pt}[\ht\strutbox][0pt]{$\check\delta_{\hat \xi} B_{ab}=(\partial_m \hat\xi_n-\partial_n \hat\xi_m)$},
which is precisely the gauge transformation of the Kalb-Ramond
field $B$. Thus, also the local symmetries  map correctly
under \eqref{redefine}.

Finally, employing \eqref{redefine} we can translate the $\alpha'$-corrections to \eqref{stringaction}  into the non-geometric frame. 
This provides an expansion in the derivative $D^a$, and thus  \eqref{finalaction} is a valid approximation for solutions with large radii  $\fa g^{ab}\sim \fa r\gg 1$. 
At second order,  this expansion reads
\begin{align}
  \nonumber
   &\fa S^{(1)} =  \frac{1}{2\kappa^2}
    \,\frac{\alpha'}{4}
  \int d^{26}x\, \sqrt{-|\fa g|}\, \bigl|\fa\beta^{-1}\bigr|\, e^{-2\phi}
   \Bigl( \fa R^{abcd}\, \fa R_{abcd} 
     -{\textstyle {1\over 2}}  \fa R^{abcd}\, \fa \Theta_{abm}
     \fa\Theta_{cd}{}^m \\
      & \hspace{150pt}
      + {\textstyle{1\over 24}}   \fa \Theta_{abc} \fa \Theta^{a}{}_{mn} \fa \Theta^{bm}{}_p
        \fa \Theta^{cnp} 
    -{\textstyle {1\over 8}} \hspace{1pt} (\fa\Theta^2)_{ab}  (\fa\Theta^2)^{ab}     \Bigr) \,.
\end{align}


\section{Conclusions}

We close  with some comments on open questions and future directions.
It would be interesting to study
solutions to the equations of motion \eqref{eom}
of the novel symplectic gravity  action.
In particular, we expect
analogues of the elementary string and the
solitonic five-brane solution.
It would also be interesting to compute 
next to leading order terms in the action and 
to include space-time fermions, as well as to
study the up-lift of symplectic gravity to M-theory.

The presence of a dynamical  (quasi-)Poisson structure and 
the appearance of the Seiberg-Witten limit  in relating
the non-geometric frame to the geometric one suggests
that it might be possible to perform a deformation
quantization of the classical symplectic gravity  action.
If that is feasible, we expect the non-associative structures 
observed in \cite{Blumenhagen:2010hj,Lust:2010iy} 
to play an essential role.


\vspace{4pt}

\paragraph*{Acknowledgments}
We thank Dieter L\"ust for  discussions. R.B. and F.R. thank 
the Simons Center for Geometry and Physics for hospitality.
E.P. was partially supported 
by the Netherlands Organization for Scientific Research (NWO) under the VICI grant 680-47-603,
as well as by the Padova University Project CPDA105015/10
and by the MIUR-FIRB grant RBFR10QS5J.



\end{document}